\documentclass[aps,prl,twocolumn,showpacs,superscriptaddress]{revtex4-1}
\usepackage{graphicx}
\usepackage{amsmath}
\usepackage{amssymb}
\usepackage{colordvi}
\usepackage{mathrsfs}
\usepackage{bm}
\usepackage{verbatim}
\usepackage{dcolumn}
\usepackage{bm}
\usepackage{epsfig}
\usepackage{subfigure}
\usepackage{makecell}
\usepackage[colorlinks,allcolors=blue]{hyperref}

\begin{document}
\title{Layer Zero-Line Modes in Antiferromagnetic Topological Insulators}

\author{Wenhao Liang}
\affiliation{CAS Key Laboratory of Strongly-Coupled Quantum Matter Physics, and Department of Physics, University of Science and Technology of China, Hefei, Anhui 230026, China}
\author{Tao Hou}
\affiliation{CAS Key Laboratory of Strongly-Coupled Quantum Matter Physics, and Department of Physics, University of Science and Technology of China, Hefei, Anhui 230026, China}
\author{Junjie Zeng}
\affiliation{CAS Key Laboratory of Strongly-Coupled Quantum Matter Physics, and Department of Physics, University of Science and Technology of China, Hefei, Anhui 230026, China}
\author{Zheng Liu}
\affiliation{CAS Key Laboratory of Strongly-Coupled Quantum Matter Physics, and Department of Physics, University of Science and Technology of China, Hefei, Anhui 230026, China}
\author{Yulei Han}
\email[Correspondence author:~~]{hanyulei@ustc.edu.cn}
\affiliation{Department of Physics, Fuzhou University, Fuzhou, Fujian 350108, China}
\author{Zhenhua Qiao} \email[Correspondence author:~~]{qiao@ustc.edu.cn}
\affiliation{CAS Key Laboratory of Strongly-Coupled Quantum Matter Physics, and Department of Physics, University of Science and Technology of China, Hefei, Anhui 230026, China}
\affiliation{ICQD, Hefei National Research Center for Physical Sciences at the Microscale, University of Science and Technology of China, Hefei, Anhui 230026, China}
\date{\today{}}

\begin{abstract}
  Recently, the magnetic domain walls have been experimentally observed in antiferromagnetic topological insulators MnBi$_2$Te$_4$, where we find that the topological zero-line modes (ZLMs) appear along the domain walls. Here, we theoretically demonstrate that these ZLMs are layer-dependent in MnBi$_2$Te$_4$ multilayers. For domain walls with out-of-plane ferromagnetism, we find that ZLMs are equally distributed in the odd-number layers. When domain walls possess in-plane magnetization, the ZLMs can also exist in even-number layers due to in-plane mirror-symmetry breaking. Moreover, the conductive channels are mainly distributed in the outermost layers with increasing layer thickness. Our findings lay out a strategy in manipulating ZLMs and also can be utilized to distinguish the corresponding magnetic structures.
\end{abstract}

\maketitle

In two-dimensional systems, topological edge states usually exist at the boundaries of two regimes with different topologies~\cite{Hatsugai1993PRL,Graf and Porta2013}. One is located at the topologically nontrivial/trivial boundaries in the quantum version of various Hall effects~\cite{Klitzing1980,xue2013,qiao2010,KaneMele2005,zhang2006,QiaoQVH2011}. The other, also known as ZLMs, is located at the boundaries of two different topologically nontrivial systems, e.g., quantum valley Hall insulators with different valley Chern numbers~\cite{ZLM-qiao-nanolett-2011}, quantum anomalous/valley Hall systems~\cite{pan2015}, quantum spin/valley Hall systems~\cite{wang2016}, and quantum anomalous Hall insulators with different Chern numbers~\cite{ZLM-ren-2017}, which have been widely explored in graphene-like systems~\cite{ZLM-qiao-nanolett-2011,pan2015,ZLM-ren-2017,ZLM-qiao-partition-2014,ZLM-wangke,ZLM-han,ZLM1,ZLM2,ZLM3,ZLM4,ZLM5,ZLM6,ZLM7,ZLM8,ZLM9,ZLM10,ZLM11,ZLM12,ZLM13,ZLM14,ZLM15,ZLM16,ZLM17,ZLM18,ZLM19,ZLM20,ZLM21,ZLM22}.
Previous studies demonstrate that ZLMs can be modulated by external electric/magnetic/optical fields~\cite{ZLM16,ZLM17,ZLM18}, and the domain walls structures~\cite{ZLM19,ZLM20}. Inspired by recent experimental discovery of layer Hall effect in antiferromagnetic topological insulator~\cite{layer hall effect}, we find that the spatial degree of freedom corresponding to different layers is an efficient method to manipulate the ZLMs.

Recently, intrinsic antiferromagnetic topological insulator MnBi$_2$Te$_4$~\cite{MBT-xuyong-1,MBT-xuyong-2,MBT-yaowang,MBT-liuqihang,MBT-wangjian,MBT-hanyulei,MBT1,MBT2,MBT3,MBT4,MBT-domainwall,MBT-model-1,MBT-model-2,MBT-model-3,MBT-model-twisted} has attracted intense interest since it bridges the fields of topology, magnetism and van der Waals materials, where the neighboring ferromagnetic Mn layers are coupled in an antiparallel manner displaying A-type antiferromagnetism. The unique layer degree of freedom of this van der Waals material exhibits significant influence on topological phases, e.g., topologically trivial insulator in monolayer, Chern insulator in the odd number of layers ($n > 1$), and axion insulator in the even number of layers~\cite{MBT1}. Moreover, when the temperature is below the Neel ordering temperature, the magnetic domain walls have been experimentally observed in layered MnBi$_2$Te$_4$~\cite{MBT-domainwall}, which provides an ideal platform to investigate the effect of layer degree of freedom on ZLMs.

In this Letter, we systematically study the electronic transport properties of MnBi$_2$Te$_4$ multilayers with various magnetic domain walls. For out-of-plane magnetic domain walls, we find that the counter-propagating chiral ZLMs along domain walls and quantum anomalous Hall edge states along boundaries coexist with double degeneracy, leading to quantized conductance of $2~e^2/h$. The ZLMs and edge states are only distributed in the odd number of layers equally, displaying layer-dependent features. When the magnetization of domain walls has clockwise/anticlockwise rotation, the ZLMs are no longer equally distributed in the odd-number layers due to the in-plane mirror-symmetry breaking and can also exist in the even number layers. With increasing layer thickness, we find that the conductive channels are mainly distributed in the outermost layers.

\begin{figure*}
  \includegraphics[width=18cm,angle=0]{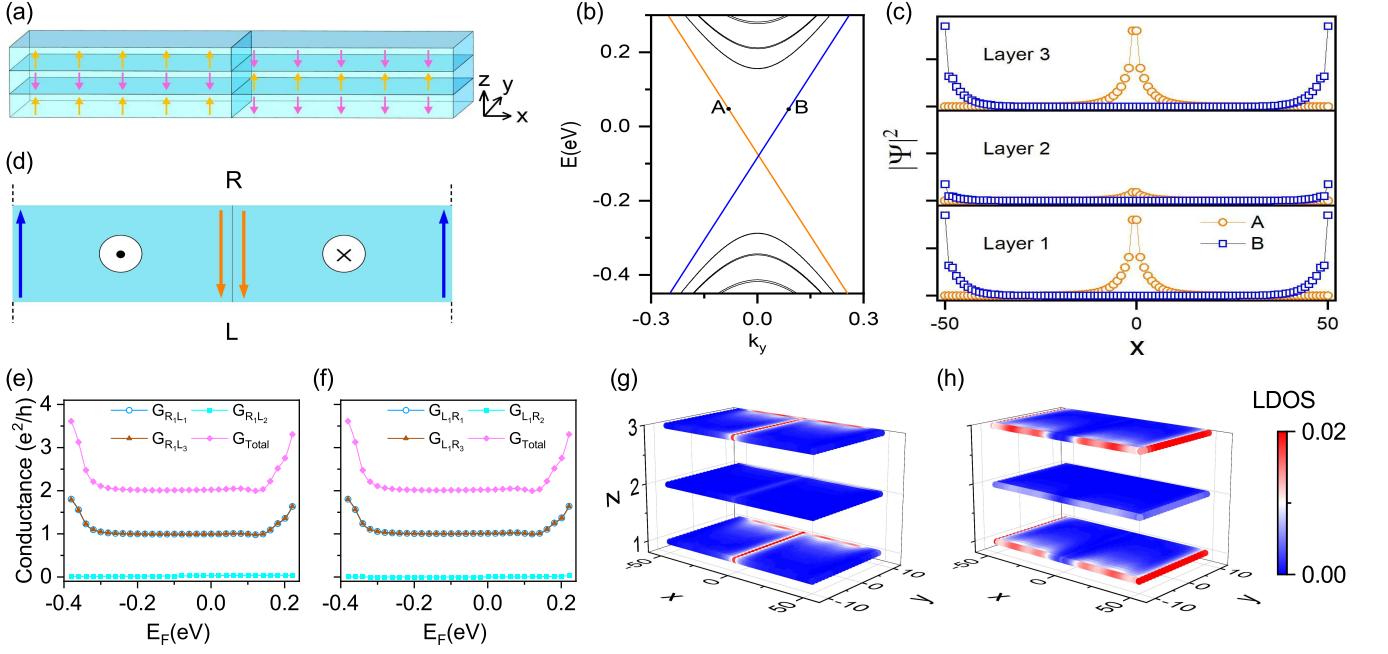}
  \caption{(a) Front view of the three-layer system; (b) Band structure of the one-dimensional nanoribbon; (c) The wavefunction distribution of states $A$ and $B$ in each layer; (d) Top view of the structure and schematic of ZLMs and edge states; (e)-(f) The four-terminal conductance from lead $R_1$ to $L_{1,2,3}$ along the $-y$ direction and from lead $L_1$ to $R_{1,2,3}$ along the $y$-direction respectively; (g)-(h) Corresponding LDOS of (e) and (f) respectively.}
  \label{fig_3layers_Mz}
\end{figure*}

The low-energy electronic properties of MnBi$_2$Te$_4$ can be described by Bi-$p_z$ and Te-$p_z$ orbitals around $\Gamma$ point~\cite{MBT-model-1}, whereas Mn-$d$ orbitals are farther away from the Fermi level. Based on the low-energy Hamiltonian of MnBi$_2$Te$_4$~\cite{MBT-model-1,MBT-model-2,MBT-model-3,MBT-model-twisted}, we construct the tight-binding model Hamiltonian~\cite{BHZ-HgTe,Bi2Se3-3D,model-jianghua} on the basis of $\left\lbrace  \left| \text{p}_\text{z,Bi}^+,\uparrow\right\rangle ,
\left| \text{p}_\text{z,Te}^-,\uparrow\right\rangle, \left| \text{p}_\text{z,Bi}^+,\downarrow\right\rangle ,
\left| \text{p}_\text{z,Te}^-,\downarrow\right\rangle \right\rbrace $:
\begin{eqnarray}\label{3D model hamiltonian}
  \nonumber
  H &=& \sum_{\left\langle ij\right\rangle \alpha}c_i^\dagger T_\alpha c_j + \sum_{i}E_0 c_i^\dagger c_i + \sum_{i}(-1)^{n_z}m_0 c_i^\dagger c_i + \text{h.c.},
\end{eqnarray}
where $c_i^\dagger$ ($c_i$) is the creation (annihilation) operator of the electron at site $i$, $\left\langle \dots \right\rangle$ denotes the nearest neighboring coupling and $\alpha=x,y,z$. The first two terms represent the bulk Hamiltonian of topological insulator, with $T_{\alpha} =(B_\alpha \sigma_0 \otimes \tau_z + D_\alpha \sigma_0 \otimes \tau_0 - \text{i}A_\alpha \sigma_\alpha \otimes \tau_x)/2$ and $E_0 = (M_0-\sum_{\alpha}B_\alpha)\sigma_0 \otimes \tau_z - \sum_{\alpha}D_\alpha \sigma_0 \otimes \tau_0$. The third term describes A-type antiferromagnetic interlayer coupling with $m_0 =m\sigma_z \otimes \tau_0$, where $\tau$ and $\sigma$ are orbital and spin Pauli matrices, respectively. Various magnetic domain walls are reflected by the magnetization $m$. Other parameters are set to be $A_\alpha=A=1.5$, $B_\alpha=B=1.0$, $D_\alpha=D=0.1$ and $M_0=0.3$~\cite{model-jianghua}, the magnetization strength is chosen to be $m=0.35$.

The electronic transport properties are evaluated by the Landauer-Buttiker formula~\cite{Landauer-Buttiker formula}:
\begin{eqnarray}
 \nonumber 
  G_{pq} = \frac{2e^2}{h}\text{Tr}[\mathrm{\Gamma}_p G^\text{r} \mathrm{\Gamma}_q G^\text{a}],
  \label{Landauer-Buttiker formula}
\end{eqnarray}
where$G^{r/a}$ are the retarded/advanced Green's functions of the central scattering region. $\mathit{\Gamma}_{p/q}=\text{i}[\mathit{\Sigma}_{p/q}^\text{r}-\mathit{\Sigma}_{q/q}^\text{a}]$ is the linewidth function describing the coupling between lead $p/q$ and the central scattering region with self-energy $\mathit{\Sigma^{r/a}}$ of the leads. The local density of states (LDOS) injected from lead $p$ is represented by $\rho_p(r,\varepsilon_\text{F})=1/{2\pi}[G^\text{r} \mathit{\Gamma}_p G^{a}]_{rr}$~\cite{Landauer-Buttiker formula}, where the Fermi energy is set to be $\varepsilon_\text{F}/B=0.135$.

We first construct a three-layer domain wall model with out-of-plane magnetization [see Fig.~\ref{fig_3layers_Mz}(a)]. The magnetic moments are represented by yellow and pink arrows in each layer. In our system, the $y$-direction is infinite, whereas the $x$- and $z$-directions are finite. Due to the quantum anomalous Hall nature of intrinsic A-type antiferromagnetic topological insulator, the left and right sides of the domain wall have opposite Chern numbers $\mathcal{C}$~\cite{qiao2010}, i.e., $\mathcal{C}=1/-1$ for the left/right side.

Figure~\ref{fig_3layers_Mz}(b) displays the one-dimensional band structure, where the bulk bands, doubly-degenerate ZLMs (\textit{A}) and edge states (\textit{B}) are respectively denoted in black, orange and blue. The wavefunction distributions of states \textit{A} and \textit{B} are different, i.e., as shown in Fig.~\ref{fig_3layers_Mz}(c), the \textit{A} states are dominantly distributed in the domain wall (middle) region, whereas the \textit{B} states are mainly distributed at the boundaries. We can also find that both ZLMs and edge states are mainly located at the bottom and top layers with the same chirality of magnetization. These states are suppressed with backscattering and robust against weak disorders due to the spatially separated counter-propagation channels localized at interfaces and boundaries, respectively~\cite{ZLM-wangke}. The left(right) side of the sample possesses net upward(downward) magnetization as shown by the black dot(cross) in the circle in Fig.~\ref{fig_3layers_Mz}(d). The orange arrows in the middle indicate the ZLMs, whereas the blue ones at the boundaries are edge states. The propagation directions of ZLMs and edge states are opposite.

To study the electronic transport properties, we consider a four-terminal mesoscopic device, where the electron injecting from one layer transits through the three-layer device and then exits from three independent terminals connected respectively with the three layers. Due to the presence of in-plane mirror symmetry, the bottom and top layers are equivalent. Thus, we focus on two situations, i.e., the incident terminal is linked with the bottom and middle layers, respectively.

Figures~\ref{fig_3layers_Mz}(e)-\ref{fig_3layers_Mz}(h) display the four-terminal conductances as a function of Fermi energy and the corresponding LDOS. The subscript $i$ of $L$($R$) represents that the lead is the $i$th layer at $L$($R$) side. $R_{i}L_{j}$($L_{j}R_{i}$) represents that the electronic transport is from lead $R_{i}$($L_{j}$) to $L_{j}$($R_{i}$). When the electron is injected from $R_1$ to $L_{1,2,3}$ [see Fig.~\ref{fig_3layers_Mz}(e)], the conductance $G_{R_{1}L_{1}}$ and $G_{R_{1}L_{3}}$ are quantized to $e^2/h$, whereas $G_{R_{1}L_{2}}$ vanishes, and the total conductance is quantized to be $2~e^2/h$ when the Fermi energy is within the bulk gap, indicating that the electron is equally distributed in the layers with the same magnetization direction via interlayer coupling. When the electron is injected from $L_1$ to $R_{1,2,3}$ [see Fig.~\ref{fig_3layers_Mz}(f)], the conductance is the same as that from $R_1$ to $L_{1,2,3}$, but the physical origins are different, i.e., the electrons from sides $R$ to $L$ are mainly distributed around the domain walls in the bottom and top layers [see Fig.~\ref{fig_3layers_Mz}(g)], while the electrons from sides $L$ to $R$ are mainly located at the boundaries in the odd-number layers [see Fig.~\ref{fig_3layers_Mz}(h)], indicating that $G_{R_1L_{1,2,3}}$ and $G_{L_1R_{1,2,3}}$ originate from the ZLMs and the edge states, respectively. When the electron is injected from the middle layer, the corresponding conductances of three independent outgoing terminals become vanishing, indicating the insulating nature of the middle layer.

\begin{figure}
  \centering
  \includegraphics[width=8cm,angle=0]{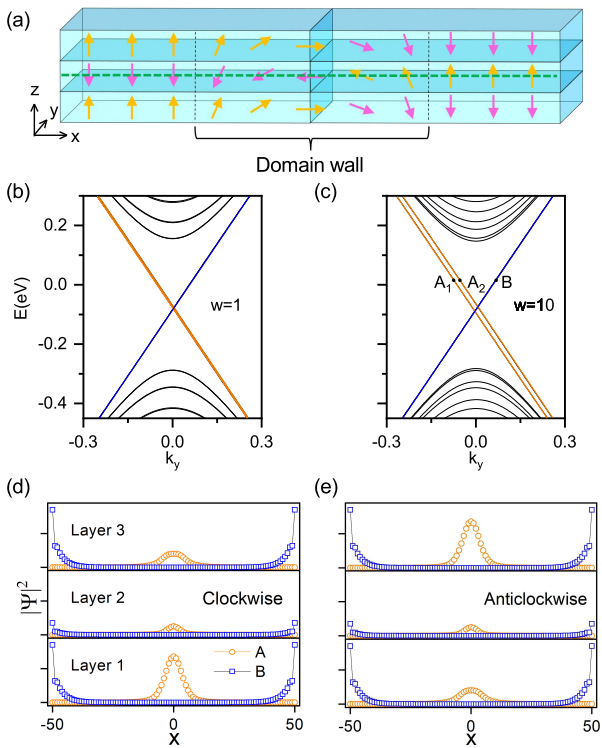}
  \caption{(a): The three-layer systems with in-plane magnetization. The green dashed line in the middle layer is the mirror of clockwise configuration related to anticlockwise configuration; (b)-(c): Band structures with different domain wall widths represented by w; (d)-(e): The wavefunction distribution of $A_1$, $A_2$, and $B$ states in every layer respectively, where $A$ is the total of $A_1$ and $A_2$.}
  \label{fig_3layers_C-AC}
\end{figure}

Besides above domain wall configuration with out-of-plane magnetization, the magnetic moments may be deviated from the out-of-plane direction, leading to in-plane magnetic moments as experimentally observed~\cite{MBT-domainwall}. Two kinds domain wall configurations with in-plane magnetization are considered, i.e., clockwise configuration displayed in Fig.~\ref{fig_3layers_C-AC}(a), and anticlockwise configuration related to clockwise configuration by in-plane mirror symmetry (the green dashed line in the middle layer is the mirror).
Unlike the domain wall structure with out-of-plane magnetization, the in-plane magnetization gradually flips in each layer in the domain wall region leading to finite domain wall width, which can further influence the electronic properties of ZLMs. Figures~\ref{fig_3layers_C-AC}(b) and \ref{fig_3layers_C-AC}(c) plot the one-dimensional band structures of the in-plane domain wall systems with different domain wall widths.
For a narrow domain wall width [Fig.~\ref{fig_3layers_C-AC}(b)], the edge states (blue) keep degenerate whereas the ZLMs (orange) start to split in energy. Along with the broadening of domain wall width [Fig.~\ref{fig_3layers_C-AC}(c)], the energy splitting of ZLMs enlarges whereas the edge states remain unchanged. Although the configurations of in-plane magnetization do not affect the band structures and edge states, they have influence on the wavefunction distributions of ZLMs. In the clockwise arrangement, as shown in Fig.~\ref{fig_3layers_C-AC}(d), the ZLMs (labeled as $A$) are dominantly distributed in the domain wall (middle) region with a larger amplitude in the bottom layer; whereas in anticlockwise configuration, as displayed in Fig.~\ref{fig_3layers_C-AC}(e), the ZLMs are mainly distributed in the top layer of the domain wall region. This magnetic orientation-dependent topological conducting states can be used as current beam splitters in low-energy consumption electronics.

\begin{figure}
  \centering
  \includegraphics[width=8.0cm,angle=0]{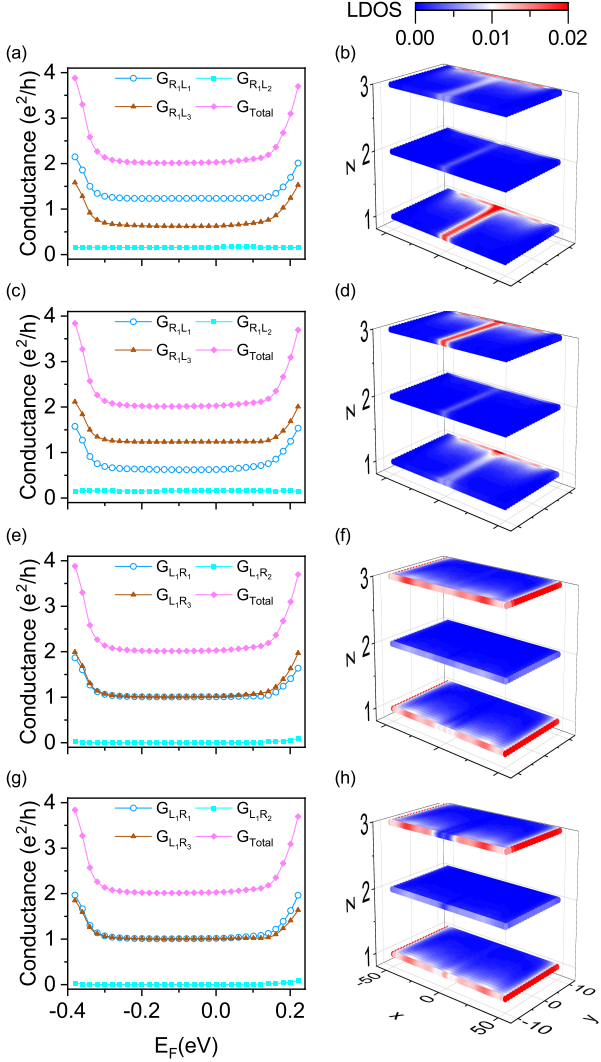}
  \caption{The four-terminal conductances and corresponding LDOS with in-plane magnetization. (a)-(d): From lead $R_1$ to $L_{1,2,3}$ along $-y$-direction, where (a)-(b) are for the clockwise configuration and (c)-(d) are for the anticlockwise configuration; (e)-(h): From lead $L_1$ to $R_{1,2,3}$ along $y$-direction, where (e)-(f) are for the clockwise configuration and (g)-(h) are for the anticlockwise configuration.}
  \label{3layers_4-terminal_C-AC}
\end{figure}

To further explore the electronic properties, we consider three different situations, i.e., the current coming from the three independent layer, respectively. Although the top and bottom layers are nonequivalent due to the breaking of in-plane mirror symmetry, we find that the four-terminal conductances and the corresponding LDOS are the same for the current incoming from the top or bottom layer. For the current incoming from the middle layer, the conductances of three independent terminals are vanishing, indicating the insulating nature of the system. Hereinbelow, we focus on the system with current incoming from the bottom layer.

Figure~\ref{3layers_4-terminal_C-AC} displays the four-terminal conductances and corresponding LDOS with different in-plane magnetization configurations. As aforementioned, the conductance from $R$ to $L$ side is originated from the ZLMs [see Figs.~\ref{3layers_4-terminal_C-AC}(a)-\ref{3layers_4-terminal_C-AC}(d)], whereas the conductance from $L$ to $R$ side comes from the edge states [see Figs.~\ref{3layers_4-terminal_C-AC}(e)-\ref{3layers_4-terminal_C-AC}(h)], with the total conductance being quantized to $2e^2/h$ for Fermi level inside the bulk gap.

For the clockwise magnetization [see Figs.~\ref{3layers_4-terminal_C-AC}(a)-\ref{3layers_4-terminal_C-AC}(b)], $G_{R_{1}L_{1}}\approx 1.24~e^2/h$ is about twice of $G_{R_{1}L_{3}}$, and $G_{R_{1}L_{2}}\approx 0.15~e^2/h$. For the anticlockwise magnetization, as shown in Figs.~\ref{3layers_4-terminal_C-AC}(c)-\ref{3layers_4-terminal_C-AC}(d), the conductances $G_{R_{1}L_{3}}$ and $G_{R_{1}L_{1}}$ are reversed, and $G_{R_{1}L_{2}}\approx 0.15~e^2/h$ keeps unchanged for Fermi levels inside the bulk gap. In contrast, the edge states induced conductance $G_{L_{1}R_{1}}$ and $G_{L_{1}R_{3}}$ are quantized to be $1.0~e^2/h$, whereas $G_{L_{1}R_{2}}$ is vanishing, which are independent of the magnetization configurations [see Figs.~\ref{3layers_4-terminal_C-AC}(e) and \ref{3layers_4-terminal_C-AC}(g)]. One can also observe that the edge states are distributed at the boundaries of the top and bottom layers from the LDOS [see Fig.~\ref{3layers_4-terminal_C-AC}(f) and ~\ref{3layers_4-terminal_C-AC}(h)]. The different distributions of ZLMs at the bottom and top layers can be attributed to the mirror symmetry breaking from the in-plane magnetization. In fact, the bottom (top) layer of the clockwise configuration is related to the top (bottom) layer of the anticlockwise configuration by in-plane mirror symmetry. The above results indicate that the conductance and current distributions can be easily tuned by the in-plane magnetization configurations and the direction of incident lead.

We also explore the electronic properties of a five-layer system in the Supplementary Materials~\cite{Supplementary Material}. Similar to the results of the three-layer system, we find that the total conductance for Fermi level inside the bulk gap is always quantized to $2.0~e^2/h$, regardless of the number of layers. In the presence of only out-of-plane magnetization, the ZLMs are still mainly distributed at the odd number layers, with a larger distribution at the outermost layers, whereas the wavefunction of edge states is also distributed at the odd number layers with equal amplitude. In the presence of in-plane magnetization, although the distribution of edge states is unchanged, the distribution of ZLMs can be effectively tuned by the clockwise/anticlockwise magnetization.

\begin{figure}
  \centering
  \includegraphics[width=8.6cm]{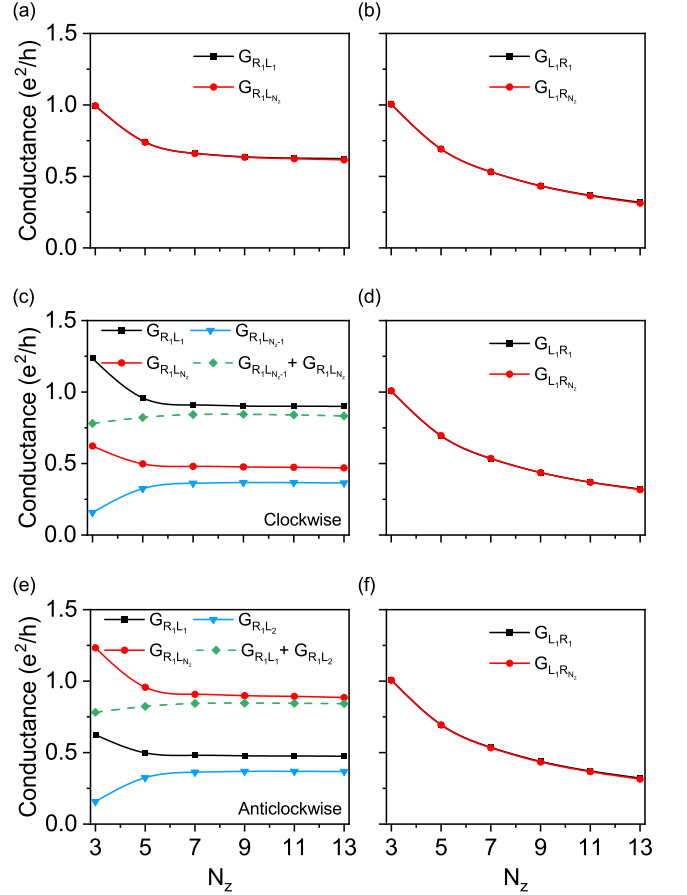}
  \caption{Thickness dependence of conductance in multilayer domain wall system. (a)-(b) For the system with purely out-of-plane magnetization; (c)-(d) For the system with clockwise in-plane magnetization; (e)-(f) For the system with anticlockwise in-plane magnetization. The conductance in the left column is contributed from ZLMs whereas that in right column originates from the edge states.}\label{trans_multilayers}
\end{figure}

The above electronic transport characteristics are also applicable to thicker samples with an antiferromagnetic domain walls. To quantitatively investigate the thickness dependence of electronic distribution in multilayers, we calculate the electronic transport properties of the multi-terminal system connected separately with each lead, i.e., the input lead is only connected with the bottom layer whereas the output leads connect separately with each layer of the system.

Figures~\ref{trans_multilayers}(a) and \ref{trans_multilayers}(b) display the conductances as a function of the number of layers $N_z$ with only the out-of-plane magnetization. Because the wavefunction is mainly distributed at the outermost layers for out-of-plane magnetization, we focus on the conductance of the corresponding layers, $G_{R_{1}L_{1}}$ and $G_{R_{1}L_{N_z}}$. As shown in Fig.~\ref{trans_multilayers}(a), the conductances $G_{R_{1}L_{1}}$ and $G_{R_{1}L_{N_z}}$ from ZLMs are equal, and gradually decay from $1.0 ~e^2/h$ for $N_z=3$ to $0.66 ~e^2/h$ for $N_z=7$, then decrease slowly when $N_z>7$, indicating that the conductance leaving from the outermost layers are gradually saturated with the increase of the layer number. It is noteworthy that the total conductance is quantized to $2.0 ~e^2/h$ for Fermi energies inside the bulk gap, implying that $G_{R_{1}L_{1}}$ and $G_{R_{1}L_{N_z}}$ are still dominant for the electronic transport properties. Since the edge states are distributed at the odd number layers with equal amplitude, the conductance $G_{L_{1}R_{1}}$/$G_{L_{1}R_{N_z}}$ from quantum anomalous Hall edge states gradually decreases to $\frac{2}{n} ~e^2/h~(N_z=2n-1)$ with the increase of thickness, where $n$ denotes the number of odd layers [see Fig.~\ref{trans_multilayers}(b)]. The conductance from edge states as a function of thickness is the same for both out-of-plane and in-plane magnetizations [see the right column of Fig.~\ref{trans_multilayers}].

In the presence of clockwise in-plane magnetization, as displayed in Fig.~\ref{trans_multilayers}(c), $G_{R_{1}L_{1}}$ and $G_{R_{1}L_{N_z}}$ from ZLMs are different, i.e., $G_{R_{1}L_{1}} > G_{R_{1}L_{N_z}}$. However, their variations are similar, i.e., they gradually decrease from $N_z=3$ to $N_z=7$, then become almost unchanged for $Nz>7$. The presence of clockwise in-plane magnetization leads to wavefunction distribution in the even number of layers (e.g., $N_z-1$), indicating a conducting channel (nonzero conductance of $G_{R_{1}L_{N_z-1}}$). We can find that $G_{R_{1}L_{N_z-1}}$ gradually increases as a function of layer thickness with a saturated value of $0.36 ~e^2/h$.
The sum of $G_{R_{1}L_{N_z}}$ and $G_{R_{1}L_{N_z-1}}$ are about $0.84 ~e^2/h$ and are comparable with $G_{R_{1}L_{1}}$ [see the green dashed line in Fig.~\ref{trans_multilayers}(c)], which implies that the three conductances dominate the electronic transport properties.
In the presence of anticlockwise magnetization, the results are similar to that of clockwise magnetization, but with a relation of $G_{R_{1}L_{N_z}}> G_{R_{1}L_{1}}>G_{R_{1}L_{2}}$.

In conclusion, we theoretically propose a feasible scheme to achieve topologically protected conducting channels in the realistic MnBi$_2$Te$_4$ systems with different types of magnetic domain walls, and the distributions of these channels are layer-dependent and can be tuned. We can use these electronic transport results to predict or distinguish the internal magnetic domain wall structures. Our study can shed light on designing low-power topological quantum devices and beam splitters, and has broad application prospects in magnetic memory and spintronic devices~\cite{MBT-magnetic memory}. Our results are not only applicable to the MnBi$_2$Te$_4$ system. When the system becomes layer-antiferromagnetic topological insulators with domain walls, the same layer-dependent electronic transport properties can be obtained.

This work was financially supported by the NNSFC (No.11974327), Fundamental Research Funds for the Central Universities (WK3510000010, WK2030020032), Anhui Initiative in Quantum Information Technologies (AHY170000). We also thank the Supercomputing Center of University of Science and Technology of China for providing the high performance computing resources.

\end{document}